\documentstyle[12pt]{article}
\begin{document}

\bigskip

\centerline {\bf {Vertical scaleheights in a gravitationally coupled,}}
\centerline {\bf {three-component Galactic Disk}}
\medskip

\bigskip

\bigskip

\centerline{Chaitra A. Narayan and Chanda J. Jog}
\centerline {Department of Physics, Indian Institute of Science}
\centerline { Bangalore 560 012, INDIA.}
\centerline {email: chaitra@physics.iisc.ernet.in,cjjog@physics.iisc.ernet.in}

\bigskip

\bigskip

\centerline{Astron. \& Astrophys. 2002, 394, 89.}

\newpage

\noindent {\bf Abstract}

\noindent The vertical scaleheight of the atomic hydrogen gas shows a 
remarkably flat distribution with the galactocentric radius in the inner Galaxy.
This has been a long-standing puzzle (Oort 1962) because the gas 
scaleheight should increase  with radius when treated as 
responding to the gravitational potential of the exponential stellar disk.
We argue that the gravitational force of the molecular and atomic 
hydrogen gas should also be brought into the picture to explain this.
We treat the stars, the HI and H$_2$ gas as three gravitationally coupled 
components in the Galactic disk, and find the response of each component
to the joint potential and thus obtain their vertical distribution in a 
self-consistent fashion.
The effect of the joint potential is different for the three components
because of their different velocity dispersions.
We show that this approach cohesively and naturally explains the
observed scaleheight distribution of all the three components,
namely, the HI and H$_2$ gas and the stars,
 in the region studied (2-12 kpc).
This includes the constant scaleheight for the HI seen in the inner Galaxy. 
The effect of H$_2$ dominates in the molecular ring region
of 4-8.5 kpc, while that due to HI is dominant in the outer Galaxy.  

\noindent Running Title : Scaleheights of Stars and Gas in the Galaxy

\bigskip

\noindent {\bf Key Words:} galaxies: ISM - galaxies: kinematics and 
dynamics - galaxies: structure - galaxies: HI - galaxies: The Galaxy
- hydrodynamics

\newpage

\noindent {\bf 1. Introdution}

The constancy of HI vertical scaleheight in the inner region 
of our Galaxy ($< 8.5$kpc) has been well-known for a long time
and has not been explained so far (Oort 1962, Dickey \& Lockman 1990,
Heiles 1991). This
behaviour is surprising since the atomic hydrogen gas, in the presence of the 
stellar disk potential alone, should have a scaleheight which increases 
exponentially with radius. Physically, the scaleheight of a component is a 
measure of the equilibrium between the local vertical gravitational
force and the gas pressure  as given by the equation of
hydrostatic equilibrium  (e.g., Rohlfs 1977). Thus, an increase in the 
gravitational force in the disk would reduce the scaleheight. We show in 
this paper that the gas gravity needs to be taken into account, to get 
the correct physical description for the observed vertical scaleheights 
of all the disk components.

The interstellar gas in the Galaxy contains  
$\sim$ 15\% of the total disk surface density (Binney \& Merrifield 1998).
About half of it is in the form of atomic hydrogen and the other half is in 
the form of molecular hydrogen but with widely different radial distributions
(Scoville \& Sanders 1987, Bronfman et al. 1988). A large fraction of mass 
of atomic hydrogen is 
located in the outer Galaxy (with R, the galactocentric radius $ > 8.5$ kpc), 
which is also the
region where the force due to the stellar disk becomes weak. 
Hence we expect the gravity of atomic gas to play a 
significant role in the determination of the scaleheights of stars and gas in 
the outer Galaxy. In contrast, most ($\sim 80$\%) of the molecular hydrogen gas 
is concentrated in the form of a ring between $4-8.5$ kpc.
The molecular hydrogen gas is known to exist in the form of self-gravitating 
clumps called molecular clouds and several such clouds segregate to 
form a cloud complex (Rivolo, Solomon, \& Sanders 1986). 
It has been shown recently that such complexes (of a 
few $100$ pc in size each) with mass densities $\sim 6$ times that 
of Oort limit dominate the local gravitational field, and
this leads to
 a redistribution of the nearby disk matter resulting in smaller 
scaleheights of the disk components (see Jog \& Narayan 2001). 
On a larger scale, the average H$_2$ distribution will affect the 
scaleheight distribution of all the disk components in the inner Galaxy.
Because of its low velocity dispersion the gas forms a thin layer,
and hence can dominate the in-plane dynamics and affect the net
vertical distribution of the disk components even though its contribution
to the total surface density is small.

In this paper, we treat the stars, the HI and H$_2$ gas as three gravitationally 
coupled disk components and obtain their vertical scaleheights as a
function of radius under the new joint potential.
A similar study showing the importance of gravitational
coupling between stars and gas for the local stability of a 
two-component galactic disk has been shown earlier by Jog \&
Solomon (1984) and Jog (1996). The importance of including the HI self-gravity
was pointed out in the past to mainly study the vertical
distribution of HI at large radii in galaxies (van der Kruit 1988, 
Olling 1995). However, these earlier papers
do not include H$_2$ gas and also they do not treat a coupled
three-component disk as we do in this paper.

The formulation of equations is discussed in Sect. 2. Section~3 
describes the method of solving them and the parameters used. 
The results obtained and a comparison with observations are discussed 
in Sect. 4. The discussion and conclusions follow in Sect. 5 and 
Sect. 6  respectively. 

\noindent {\bf 2. Formulation of Equations} 

We consider the atomic and molecular hydrogen gas layers to be very thin 
disks embedded in the stellar disk. We use the galactic
cylindrical co-ordinates ($R, \phi, z$), and consider their distribution
from $R = 2-12$ kpc.
For the sake of simplicity, all the three disks are taken to be
axisymmetric and coplanar. 
The gravitational force due to these embedded layers 
would modify the steady-state density distribution of all the three 
components and along all the three axes.
However, we can neglect the effect along the azimuthal direction
because of the assumed axisymmetry and along the radial
direction because the disk is thin. Therefore, we need to
consider the modification of the steady-state density
distribution only along the $z$-axis.

The force equation along the $z$-axis or the equation of hydrostatic 
equilibrium is given by (e.g., Rohlfs 1977):

$$ \mathrm {\frac {\langle{({\mathit v}_z)^2_i}\rangle}{\rho_i} \: \frac {d {\rho_i}}
   {d{\mathit z}} \: = \: {({\mathit K}_z)_s} \: + \: {({\mathit K}_z)_{HI}} 
   \: + \: {({\mathit K}_z)_{H_2}} \: + \: {({\mathit K}_z)_{DM}} } \eqno (1) $$    

\noindent where $\rho$ is the mass
density, $(K_{\mathrm z}) \: = \: -  {{\partial}{\psi}}/ {{\partial}z} $ is 
the force per unit mass along $z$-axis, $\psi$ is the corresponding
potential, and the subscript {\it i} = s, HI, and H$_2$ denotes these 
quantities for stars, HI and H$_2$ respectively. The last term on
the right hand side denotes the force along the $z$-axis due to the dark 
matter (DM) halo.
Due to the disk being thin, its effect on the vertical distribution within
the halo can be neglected.
We take the root mean square of
the vertical velocities of a component ${\mathrm {\langle{({\mathit v}_z)^2_i}
\rangle }^\frac {1}{2}}$ 
or the random velocity dispersion at a radius $R$ and treat the component as 
being isothermal along $z$.
The right hand side of equation (1) gives the total vertical force due to all 
the components.
The dark matter halo has been included for the sake of completeness, and also
because it helps us to quantify the role played 
by the halo in defining the vertical density distribution in the region of 
interest ($R \leq 12$ kpc).

\noindent For a thin axisymmetric disk, the joint Poisson equation reduces to :

$$ {\mathrm {\frac {d^2{\psi_s}}{d{\mathit z^2}} \: + \frac {d^2{\psi_{HI}}}
   {d{\mathit z^2}} \: + \frac {d^2{\psi_{H_2}}}{d{\mathit z^2}}
   \: = \:  4 \pi {\mathit G} \left ({\rho}_s \: + \: {\rho}_{HI} \: + \: 
   {\rho}_{H_2} \right )}} \eqno (2) $$ 

\noindent Combining equations (1) and (2), the density distribution 
of a component at a radius $R$, can be defined by : 

$$ {\mathrm {\frac { d^2{\rho_i}}{d{\mathit z^2}} \: = \: \frac {\rho_i}
   {\langle({\mathit v}_z)^2_i \rangle} \: \left [ - 4 \pi {\mathit G} \: 
   ({\rho_s} + 
   {\rho_{HI}} + {\rho_{H_2}}) + \frac {d({\mathit K}_z)_{DM}}{d{\mathit z}} 
   \right ] \: + \: \frac {1}{\rho_i} \: \left ( \frac {d{\rho_i}}{d{\mathit z}} 
   \right )^{2} }} \eqno (3) $$

\noindent where the square brackets contain terms that arise due
to the joint potential of the three disk components and the halo,
and  the same total potential is experienced by all the components.
The vertical velocity dispersion, on the other hand, varies with 
each component. Thus, despite a common gravitational potential,
the density distribution of each 
component will be  different due to the difference in their random 
velocity dispersions.

\noindent {\bf 3. Solution and Parameters}

\noindent {\bf 3.1. Solution of equations}

We need to solve the three coupled equations (represented by equation (3)) 
simultaneously to obtain the vertical density 
distribution of each component.
Each second order ordinary differential equation can be split into
two first order differential equations for the sake of simplicity.
They can be solved numerically as an initial value problem, using the 
fourth-order Runge-Kutta method of integration (Press et al. 1986).
The two boundary conditions required at the mid-plane, $z = 0$  are :

$$ \mathrm {\rho_i \: = (\rho_\circ)_i \: \: and  \: \: \frac 
   {d\rho_i}{d{\mathit z}} \: = 0 } \eqno (4) $$

For a realistic distribution, the density along the vertical axis 
is homogeneous very close to the mid-plane, thus 
$ \mathrm {{d{\rho_i}}/{d{\mathit z}} = 0 \: at \: {\mathit z} = 0 }$.  
We are then left with $\mathrm {(\rho_{\circ})_i}$, the modified midplane density 
which is not known a priori. 
The distribution of matter can be treated 
as a one dimensional problem along the $z$-axis and hence the surface 
density $\Sigma_{\mathrm i}(R)$ will not vary even when the joint gravitational 
potential is considered.
The surface density is twice the area under the curve $\rho_{\mathrm i}(z)$ 
versus $z$. 
Given a value of $\Sigma_{\mathrm i}(R)$ (see Sect.~3.2 for the values used), the 
value of $\mathrm {(\rho_{\circ})_i}$ can be found by trial and error.
Once this is fixed, the distribution $\rho_{\mathrm i}(z)$ follows easily. 
        
All the three components, stars, HI and H$_{2}$ affect each other's
density distribution via equation (3) so that the Galactic disk is actually a 
coupled system.
At each $R$, the three density functions are solved simultaneously by
taking account of the effect of the other components in an iterative
fashion.  
First, $\rho_{\mathrm s}(z)$ is evaluated using equation (3) 
with null values for the corresponding gas densities. $\rho_{\mathrm {HI}}(z)$ 
is then obtained by using the known stellar density distribution and null 
values for $\rho_{\mathrm H_2}(z)$. Knowing $\rho_{\mathrm s}(z)$ and 
$\rho_{\mathrm {HI}}(z)$, $\rho_{\mathrm H_2}(z)$ can be found easily. 
However, these results do not describe the real coupled disk
distribution because $\rho_{\mathrm s}(z)$ has been evaluated here in 
the absensce of HI and H$_2$. 
Knowing the non-zero values for gas densities, $\rho_{\mathrm s}(z)$ is 
re-evaluated incorporating the gas gravity. 
The above cycle is repeated four times until each of the 
distribution converges with a fifth decimal accuracy.
We obtain a sech$^2$-like distribution for each 
component and we use its HWHM (half-width-half-maximum) to define the 
vertical scaleheight. 
In comparison, for a one-component self-gravitating disk, the vertical 
distribution obeys a sech$^2$ distribution (Spitzer 1942).
Repetition of the above calculation at regular intervals of $R$ enables us 
to plot the scaleheights versus radius, provided the surface density
$\Sigma_{\mathrm i}(R)$ is known at all radii.

\noindent{\bf 3.2. Parameters Used}

For each disk component, we need to specify the surface density 
and the random velocity dispersion at each radius $R$ in the disk.
Table 1 gives values of all the observed parameters used. 
The observed values are used for all the gas parameters, whereas
all the stellar parameters except for the velocity dispersion are
taken from models in the literature.
The HI gas surface density is negligible at the 
Galactic centre, slowly increases to 
$\sim5$ M$_{\odot}$ pc$^{-2}$ by 4 kpc and remains roughly constant 
till about $16$ kpc (Scoville \& Sanders 1987). A similar HI profile was 
obtained by Dame (1993), also based on the  HI data by Burton \& Gordon (1978).
The molecular hydrogen gas is concentrated in the form of a ring with the peak
surface density of $\sim 20$ M$_{\odot}$pc$^{-2}$ at $\sim$ 5 kpc from the 
centre (Scoville \& Sanders 1987).  
The vertical velocity dispersions of HI and H$_2$ gas are  
8 kms$^{-1}$ (Spitzer 1978) and 5 kms$^{-1}$ (Clemens 1985, and 
Stark 1984) respectively and they remain constant with radius. 
These values agree fairly well with
the determination based on a tangent-point analysis by Malhotra
(1994) for H$_2$, and by Malhotra (1995) for HI.

Lewis and Freeman (1989) have measured the stellar radial velocity 
dispersion at different points between 1-17 kpc along the 
galactocentric radius towards the Baade's window in the Milky Way.
Assuming the ratio of the vertical to the radial random velocity dispersion
at all radii to be equal to its value at the solar neighbourhood, namely $1/2$
(Binney \& Merrifield 1998), we get the corresponding 
vertical velocity dispersions. The method of
least square fit to the data gives an exponential fit with a scalelength
of 8.7 kpc, and a value of
$18$ kms$^{-1}$ at the solar neighbourhood.

The two key parameters required to find the 
entire stellar disk surface density distribution are the local stellar 
surface density and the exponential radial disk scalelength, $h_{\mathrm R}$.
The stellar disk mass surface density at the solar point 
has the following range of observed values. 
For example, from a set of distance and velocity data, Kuijken 
\& Gilmore (1991) obtain  the total disk plus halo surface 
density to be $ 48 \pm 9$ M$_{\odot}$pc$^{-2}$
for the region very close to the midplane.
More recent observations point to a value of $ 52 \pm 13$ 
M$_{\odot}$pc$^{-2}$ (Flynn \& Fuchs 1994).
Denhen and Binney (1998) use a lower limit of 
40 M$_{\odot}$pc$^{-2}$ for the total disk surface density 
as a constraint for four models involving a range of values 
for the stellar disk scalelength.

The determination of the radial disk scalelength $h_{\mathrm R}$ has  attracted 
much attention in the literature in recent years.
A wide range of values is obtained for $h_{\mathrm R}$  ranging from 
2.3 kpc (Drimmel \& Spergel 2001) to 6 kpc 
(Mendez \& van Altena 1998). 
Most of the recent papers tend towards a lower value in this range :
Fux and Martinet (1994) : 1.9-3.3 kpc ; 
Ruphy et al. (1996) : 2.3 kpc ; 
Denhen and Binney (1998) : 2-3.2 kpc ; 
Mera, Chabrier,\& Schaeffer (1998) : 3.2 kpc;
Porcel et al. (1998) : 2.1 kpc ; 
Drimmel and Spergel (2001) : 2.3 kpc.

We adopt the model of Mera et al. (1998)
(see Table 2) as {\it the standard mass model} for the Galaxy for the 
following reasons. 
First, it is modern and simple and also allows us to study the 
gravitational effect of various components.
Second, their choice of the total surface density of 
52 M$_{\odot}$pc$^{-2}$ at the solar neighbourhood from a recent paper in the 
literature (Flynn \& 
Fuchs 1994) and their choice of $h_{\mathrm R} = 3.2$ kpc fall within the 
acceptable range 
as can be seen from the discussion above.
Third, for the sake of internal consistency, we prefer to use the
parameters from a single mass model as opposed to choosing them in an ad hoc 
manner. Finally, they use a screened halo-density profile 
which is sufficient to determine the dynamical effect of the spherical halo. 
Its contribution to the local surface density is negligible but it gives rise 
to a non-zero force term along the $z$-axis.
Subtracting the total gas surface density of 7  M$_{\odot}$pc$^{-2}$ 
(Scoville \& Sanders 1987) from the above value of the local total surface 
density, we get the local stellar surface density to be 45 
M$_{\odot}$pc$^{-2}$.
This gives the central extrapolated stellar surface density, 
$\mathrm {(\Sigma_{\circ})_s}$ = 640.9  M$_{\odot}$pc$^{-2}$ as given in Table 2.

In spherical co-ordinates, the density profile for the halo is 
(Mera et al. 1998) :

$$ {\rho_{\mathrm {DM}}}(r) \: = \: \frac {v^2_{\mathrm {rot}}}{4 \pi G} \:
        \frac {1}{(R^2_{\mathrm c} + r^2)}  \eqno (5) $$

\noindent where ${\rho_{\mathrm {DM}}}$ is the dark matter halo mass density;
$R_{\mathrm c}$, the core radius = 5 kpc ; and $v_{\mathrm {rot}}$, the 
circular velocity = 220 kms$^{-1}$.

By inverting the Poisson equation for the dark matter halo, 
we calculate the halo potential to be the following :

$$  {\psi_{\mathrm {DM}}}(r) \: = \: {v^2_{\mathrm {rot}}} \: \left [ 1 \: - 
    \: \frac{1}{2} \: {\log ({R^2_{\mathrm c}} + r^2)} \: - \: \frac 
    {R_{\mathrm c}}{r} \: \mathrm {tan^{-1}} \left (\frac {r}{R_{\mathrm c}} 
    \right ) \: \right ] \eqno (6) $$

\noindent Rewriting the above equation in cylindrical co-ordinates and taking 
the second derivative of the halo potential with respect to $z$, we get

$$\frac {{\partial}^2 \psi_{\mathrm {DM}}}{{\partial} z^2} \: = \: 
  \frac {v^2_{\mathrm {rot}} R_{\mathrm c}}{(R^2 + z^2)^ {\frac{3}{2}}} 
  \: {\mathrm {tan^{-1}}} \left ( \frac {\sqrt{R^2 + z^2}} {R_{\mathrm c}} 
  \right ) \: \left [ 1 - \frac{3z^2}{R^2+z^2} \right ] \:  + \: $$
$$\: \: \: \frac {z^2 R^2_{\mathrm c} v^2_{\mathrm {rot}}}{(R^2 + z^2)^2 
  (R^2_{\mathrm c} + R^2 + z^2)} \: + \: \frac {v^2_{\mathrm {rot}}}
  {(R^2 + z^2)} \: \left [ \frac {2z^2}{(R^2 + z^2)} - 1  \right ]  \eqno (7) $$

\noindent Thus, $\mathrm {d({\mathit K}_z)_{DM}/d{\mathit z} =
 -{\partial}^2\psi_{DM}/{\partial} {\mathit z}^2}$, 
is the halo contribution used in the right hand side of equation (3).

\noindent {\bf 4. Results}

\noindent {\bf 4.1. Results for Vertical Scaleheight : Standard Model}

The results for the vertical scaleheight are obtained as a
function of the galactocentric radius using 
the Galactic mass model of Mera et al. (1998), as explained in Sect. 3.1. 
The sampling is done at every 420 pc which is set by the bin-size of 
the H$_2$ data given by Scoville \& Sanders (1987).
 
\noindent {\bf 4.1.1. HI Scaleheight}

Figure 1a shows the plot of vertical scaleheight versus radius for HI
with the dashed line obtained using the stellar potential and the solid line 
obtained using the joint potential approach. The observed values are
shown as crosses and are taken  from Lockman (1984) for the inner Galaxy 
($R < 8.5 $ kpc), and Wouterloot et al. (1990) for the outer Galaxy 
($R > 8.5 $ kpc) - see Burton (1992) for details.
The curve obtained using the stellar potential alone increases 
exponentially and thus deviates strongly from the observed 
curve beyond 8 kpc.
On using the joint potential, the scaleheights
reduce significantly at large radii and show a better overall
agreement with observations. 
Thus our model explains the old puzzle (Oort 1962) of nearly-constant scaleheight
observed in the inner Galaxy.

At 10 kpc, the scaleheight reduces by about $34\%$ to give a value of 187 pc, 
which is very close to the observed value of 193 pc
(Wouterloot et al. 1990). 
In the outer Galaxy, the HI surface density is either comparable to 
or more than that of stars because of the exponential fall-off of the 
stellar surface density. 
Thus the joint self-gravitating disk 
extends well beyond the stellar disk and the HI gravity is mainly 
responsible for the scaleheight determination in that region.
In the inner Galaxy ($R < 8.5$ kpc), the combined gravity of HI and H$_2$
is responsible for the reduced scaleheights.
Unlike the smooth dashed line, the response to the joint potential
 has many small-scale dents on it. 
This is also seen later in the plots for H$_2$ and stars.
The appearance of these dents is not due to an undersampling of data points
but rather is due to the 
gravity of H$_2$ gas which shows a non-smooth radial distribution, as seen from 
the fact that
the locations of the dents coincide with the surface density peaks
of molecular hydrogen gas.
This is analogous to the local effect of a molecular cloud complex on 
the disk shown by Jog \& Narayan (2001).  

A more subtle point is that in the range
$R = $0-5 kpc both the approaches predict lower values than 
observed, implying that some other factors must be affecting 
the scaleheight. 
There could be additional physical processes that increase the 
scaleheight in the region such as, the heating due to the bar (Binney \& 
Merrifield 1998) within the central 4 kpc. 
On the other hand, beyond 10 kpc, the predicted scaleheights are larger than 
the observed values in spite of incorporating the HI gas gravity.
Thus, the dominant role played by the HI gas gravity is still not 
sufficient to bring about a complete agreement between the two. 
One would then expect that, inclusion of the halo potential would resolve 
the disagreement between the observed and theoretical curves. 
However, Fig. 1c already includes the halo potential and it brings 
about less change in the disk density distribution than expected.
A possible reason as to why the halo contribution may not 
be very important is due to its extended $z$ distribution and this is discussed in 
detail in Sect. 5. 
Thus, the agreement of the theoretical vertical scaleheights for
HI with observations is best seen in the middle galactic
range of 5-10 kpc. A small radial variation of HI velocity dispersion
leads to a better agreement with observations over a larger radial
range as shown in Sect. 4.2.

\noindent{\bf 4.1.2. H$_2$ Scaleheight}

Figure 1b shows the plot of scaleheight vs. radius for H$_2$, with the dashed 
line obtained using the stellar potential and the solid line obtained 
using the joint potential approach. The observed scaleheight values from 
Sanders, Solomon, \& Scoville (1984) for $R < 8.5 $ kpc and Wouterloot et al. (1990)
for $R > 8.5 $ kpc are shown as crosses.
Neglecting the self-gravity of HI and H$_2$ once again yields
scaleheights much larger than the observed values. On the other
hand, the curve predicted by treating the galactic disk as a
coupled, three-component system, agrees very well with observations.  
This agreement continues further upto 14 kpc though the Fig. 1b 
shows results upto only 12 kpc.
Both the theoretical results are in good agreement with the 
observations in the inner few kpc from the Galactic centre because,
this region is entirely dominated by the stellar potential so that the 
joint potential differs very little from it.

\noindent {\bf 4.1.3. Stellar Scaleheight}
	
Figure 1c shows the vertical scaleheight curves for the stellar
disk, obtained  using the potential of stellar disk 
alone (shown as a dashed line), and that obtained  using the
joint potential (shown as a solid line). The stellar
disk potential gives an exponentially increasing curve, while
the joint potential results in a nearly flat curve.
The resulting scaleheight curve exhibits the following 
detailed behaviour. In the region of 0-5 kpc the scaleheight is almost constant 
at 300 pc.
In the middle region of 5-10 kpc it shows a linear increase with a 
slope of $\sim24$ pc kpc$^{-1}$, while beyond 10 kpc it remains a 
constant at $\sim420$ pc. Without gas gravity (the dashed line, Fig. 1c), the
stellar scaleheight in the solar neighbourhood is then 550 pc. With the gas 
gravity 
(the solid line, Fig. 1c), this comes down to a reasonable 380 pc. This agrees 
well with the local observed characteristic half thickness of 350 pc 
(Binney \& Tremaine 1987, chap. 1).
We would like to stress that the near constancy of the stellar 
scaleheight upto 5
kpc (Fig. 1c) in our model comes about naturally by incorporating the gravity
of HI and H$_2$ in a standard exponential galactic disk.

Unfortunately, these results cannot be
compared with the optically deduced scaleheights in the non-local regions of 
our own Galaxy due to the 
high optical depth in the visible band. Hence, one has to compare the trend 
in the predictions with the data from external galaxies.
van der Kruit \& Searle (1981a, b) first showed from a study of edge-on 
spirals that these exhibit a remarkably constant stellar scaleheight 
with radius. However, recent data
by de Grijs \& Peletier (1997)
show a moderate increase in the stellar scaleheight with radius, in 
agreement with the trend
shown by our results in Fig. 1c. This moderate increase is shown to be a 
general result for spiral galaxies (Narayan \& Jog 2002).

A linear increase beyond 5 kpc with a small slope of 20 pc kpc$^{-1}$ has 
been argued for by Kent, Dame, \& Fazio (1991). This is obtained from a
best fit to the near-IR data from the $\it Spacelab 2$ mission for the Galaxy. 
This is in very good  agreement with our results. A similar
conclusion, based on the COBE/DIRBE data was reached  by Drimmel \& 
Spergel (2001).
Kent et al. (1991) however do not mention whether the behaviour continues 
farther into the outer Galaxy or not. 
Their motivation for using such a slope was purely to
get the best fit to the observed data and involved no dynamics 
whereas we get it physically due to the inclusion of gas gravity in the study.

\noindent {\bf 4.2. Variation in Input Parameters}

In Sect. 4.1, we saw that the results for H$_2$ and stars using 
the joint potential approach are in very good agreement with 
the observations in the region studied (2-12 kpc). In the case of HI, the best
 agreement is limited to a small region of 5-10 kpc (see Fig. 1a).
To improve the agreement with HI observations over the entire 
region studied, the predicted scaleheights should increase in
the region 0-5 kpc and decrease in the region beyond 10 kpc.
We try to obtain this by two different approaches by varying the input 
parameters as described below.

\noindent {\bf 4.2.1. Variation in the stellar disk parameters}
 
We have used the mass model of Mera et al. (1998) in Sect. 4.1 
as a realistic mass model to bring out the importance of gas gravity.
In this section, we vary the stellar disk parameters namely 
$\mathrm {({\Sigma}_{\circ})_s}$, the central surface density, and  
$h_{\mathrm R}$, the disk scalelength, freely and study the resulting variation 
 in the HI scaleheight.
For an overall agreement, the gravitational force should be
weaker in the region below 5 kpc and stronger in the region
beyond 10 kpc.
This can be brought about by a lower value of central surface 
density along with a larger radial scalelength as compared to
the model of Mera et al. (1998).
We find that the best mathematical fit to the observed data is obtained
by using $\mathrm {({\Sigma}_{\circ})_s} = 200$ M$_{\odot}$pc$^{-2}$ and 
$h_{\mathrm R} = 6$ kpc (see Table 2).
Note that these parameters are far from the typical values of 
$\mathrm {({\Sigma}_{\circ})_s} \sim 640$ M$_{\odot}$pc$^{-2}$ 
and $h_{\mathrm R} \sim$3 kpc (see Sect. 3.2).
Also, we find that the results obtained for H$_2$ and stars in this case 
deviate to a large degree from the observed behaviour.
Thus the above attempted change in the parameters is unrealistic.
Hence this is not the correct way to explain the observed radial
variation of the HI scaleheight.
  
\noindent {\bf 4.2.2. Variation in  HI velocity dispersion}

Yet another way of improving the agreement between the scaleheight 
curve of HI and the 
observations is by varying the HI velocity dispersion
with radius instead of using a constant value of 
8 kms$^{-1}$ as done earlier in Sect. 4.1. 
We find that a simple linear variation (between $R$ = 2-12 kpc) with a 
slope of -0.8kms$^{-1}$kpc$^{-1}$ is required to obtain the least $\chi^2$ value.
The value of HI gas velocity at R = 8.5 kpc is taken to be 8 kms$^{-1}$ 
(see Table 2) and is used as a 
constraint in determining the slope. In Fig. 2a, we
plot the results for HI obtained using the joint potential
plus the variation in velocity dispersion 
(as a solid line) and the observed data as crosses (see Sect.
4.1.1 for details of observed data). The results 
agree well with observations over the entire radial range studied. On
comparing with Fig. 1a, it is clear that the variable HI
velocity dispersion leads to a better overall agreement with the
observed data.

A plausible physical mechanism to explain this varying 
HI gas dispersion could be the energy input via supernovae.
As Mckee \& Ostriker (1977) proposed, the kinetic energy of the HI
clouds is regulated by the rate of supernovae.
Hence we expect the HI velocity to increase in the 
molecular ring region where a higher star formation rate and a higher
rate of formation of supernovae is expected.
Kamphuis (1993) has shown that the increase in velocity dispersion 
at smaller radii is also observed in a number of external galaxies.

Interestingly, Oort (1962) had proposed a similar increase in the
velocity dispersion of HI at smaller radii as a possible way for 
obtaining the observed constant scaleheight, though he did not 
give a physical reason for this trend.
This idea was also proposed by de Boer (1991).
 Oort (1962) had suggested a linear 
variation in HI velocity dispersion with a slope of about $-2$ 
kms$^{-1}$kpc$^{-1}$ (varying
from 13 to 5 kms$^{-1}$ between $R$ = 4-8 kpc), where the HI 
distribution is defined by the stellar potential alone.
Note that the slope that we require is smaller and is equal 
to -0.8 kms$^{-1}$kpc$^{-1}$.
This is because we have included the effect of the gas gravity, and hence
 a smaller radial variation in HI velocity dispersion is sufficient 
to give a constant vertical scaleheight for HI.

Figure 2b contains the resulting H$_2$ scaleheight
versus radius (solid line) and the observed data for H$_2$ (see
Sect. 4.1.2 for details on observations of H$_2$). 
Figure 2c contains the resulting stellar scaleheight
versus radius (solid line). 
On varying the HI velocity dispersion the scaleheights of H$_2$ 
and stars are not affected to a noticeable extent 
(compare  Fig. 2b and Fig. 2c  respectively with Fig. 1b
and Fig. 1c), because the change in the HI velocity dispersion will not
directly affect the vertical distribution of the other components.
Thus the joint potential plus a slightly varying 
HI gas dispersion seems like a plausible physical scenario which
can self-consistently explain the scaleheight distribution of
all the three galactic disk components for realistic input parameters.

\noindent {\bf 5. Discussion}

\noindent {\bf 1.} It is interesting that the maximum impact of the different
components is seen to occur in different galactocentric radial regions.
With the stellar surface density peaking at the centre of the Galaxy, the
stellar disk alone determines the scaleheights of all the components in 
the central few kpc.
The maximum effect due to H$_2$ is seen in the region of
4-8.5 kpc, with a peak at 5 kpc. 
Finally, the maximum effect due to HI is seen only beyond 8.5 kpc,
despite the fact that the HI surface density is constant between 4-16 kpc.
This is because, the inner Galaxy is entirely dominated by stars and H$_2$ 
and it is only in the outer Galaxy that the HI becomes important.
Thus the three components seem to {\it conspire} to give a nearly constant 
scaleheight for all of them in the inner Galaxy.

\noindent {\bf 2.} We have included the dark matter halo potential  
to evaluate its contribution in reducing 
the scaleheight quantitatively.
We find that the presence of the halo reduces the HI scaleheight at 
12 kpc only by $13\%$.
This is contrary to the general expectation in the literature that the 
outer Galaxy structure is dominated and defined  by the dark halo.
This is because the dark matter and visible disk 
matter dominate at entirely different range of $z$ values.
For a self-gravitating stellar disk (eq. [1]), we find that more
than 99 percent of its matter lies within $ z \leq 1 $ kpc, within radius 
of 12 kpc.
Whereas, a standard massive spherical dark matter halo of a mass of 
$10^{12}$ M$_{\odot}$ and core radius = 5 kpc (as defined in Sect. 3.2)
 has less than $6\%$ of the entire mass in the column at 12 kpc, within the 
same $z$ limit.
Therefore, the disk matter distribution along $z$ 
is not strongly dependent on the presence of halo upto the highest
radius studied here, namely $R = 12$ kpc.

\noindent {\bf 3.} We have only considered the turbulent gas
pressure associated with the vertical velocity dispersion of HI
as being responsible for its vertical support. We have not
included the pressure support due to magnetic fields and cosmic rays.
This is for simplicity, and also because these may not be
important for supporting neutral hydrogen as argued by Lockman \&
Gehman (1991). It is interesting that our resulting scaleheights
for  a three-component, gravitationally coupled galactic disk using the support
of turbulent gas pressure alone match well with observations,
this confirms the argument of Lockman \& Gehman (1991).

\noindent {\bf 6. Conclusions}

In this paper we show that the gas gravity is crucially important in the 
determination of the vertical scaleheights of all the disk components in a 
galactic disk. We treat
the galactic disk as a gravitationally coupled, three-component system
consisting of stars, atomic gas and molecular gas, and also include the 
effect of the dark matter halo. The model developed is general and is 
applied to the Galaxy in this paper. We
obtain the self-consistent vertical distribution for each
component responding to the joint potential for a radial region
of 2-12 kpc. Our approach leads naturally to a better agreement
with observations of all the three components:

\noindent {\bf 1.} The radial variation of the HI vertical scaleheight matches
fairly well with observations and the best agreement is seen between 
$R$ = 5-10 kpc. The inclusion of gas gravity can explain the 40-year old 
puzzle of the observed nearly-constant HI scaleheight.

The overall agreement 
over the entire region studied is even better if a small linear variation
with radius in the HI velocity dispersion with a slope of -0.8
kms$^{-1}$kpc$^{-1}$ between 2-12 kpc is introduced. The physical justification
for this increase at smaller radii is the higher expected supernovae rate 
in the inner Galaxy.

\noindent {\bf 2.} The radial variation of H$_2$ scaleheight obtained matches 
very well with observations upto a radius of 14 kpc. Our model gives the physical 
origin of the H$_2$ vertical scaleheight distribution, which has not been 
studied in the literature so far. 

\noindent {\bf 3.} The stellar scaleheight is found to be nearly constant 
with radius at $\sim 300$ pc in the central region of 0-5 kpc of our 
Galaxy and shows a slow linear increase beyond 5 kpc. 
This agrees well with the result  obtained by fitting the near-IR data in the 
Galaxy by Kent et al. (1991).

We have applied the above general model to two
external galxies, NGC 891 and NGC 4565 , and we find that these also show a 
similar moderate flaring with radius (Narayan \& Jog 2002).

\bigskip

\noindent {\it {Acknowledgements}}
We would like to thank the anonymous referee for useful comments.

\bigskip

\noindent {\bf {References}}

\noindent Binney, J., \& Merrifield, M. 1998, Galactic Astronomy 
(Princeton: Princeton Univ. Press)

\noindent Binney, J., \& Tremaine, S. 1987, Galactic Dynamics
(Princeton: Princeton Univ. Press)

\noindent Bronfman, L. et al. 1988, ApJ, 324, 248

\noindent Burton, W.B. 1992, in The Galactic Interstellar
  Medium, Saas-Fee Advanced course 21, Eds., 
  Pfenniger, D., \& Bartholdi, P. (Berlin: Springer)	

\noindent Burton W.B. \& Gordon, M.A. 1978, A\&A, 63, 7

\noindent Clemens, D.P. 1985, ApJ, 295, 422

\noindent Dame, T.M. 1993, in Back to the Galaxy (AIP Conf 278), 
 eds. S.S. Holt \& F. Verter (New York: AIP), 267

\noindent de Boer, H. 1991, in The Interstellar Disk-halo
  Connection in Galaxies, IAU Symp. 144, ed. H. Bloemen 
  (Dordrecht: Kluwer), 333

\noindent de Grijs, R., \& Peletier, R.F. 1997, A\&A, 320, L21

\noindent Dehnen, W., \& Binney, J. 1998, MNRAS, 294, 429

\noindent Dickey, J.M., \& Lockman, F.J. 1990, ARA\&A, 28, 215

\noindent Drimmel, R., \& Spergel, D.N. 2001, ApJ, 556, 181

\noindent Flynn, C., \& Fuchs, B. 1994, MNRAS, 270, 471

\noindent Fux, R., \& Martinet, L. 1994, A \& A, 287, L21

\noindent Heiles, C. 1991, in The Interstellar Disk-halo
 Connection in Galaxies, IAU Symp. 144, ed. H. Bloemen 
 (Dordrecht: Kluwer), 433

\noindent Jog, C.J. 1996, MNRAS, 278, 209
 
\noindent Jog, C.J., \& Narayan, C.A. 2001, MNRAS, 327, 1021 

\noindent Jog, C.J., \& Solomon, P.M. 1984, ApJ, 276, 114

\noindent Kamphuis, J.J. 1993, PhD Thesis, University of Groningen

\noindent Kent, S.M., Dame, T.M., \& Fazio, G. 1991, ApJ, 378, 131

\noindent Kuijken, K., \& Gilmore, G. 1991, ApJ, 367, L9

\noindent Lewis, J.R., \& Freeman, K.C. 1989, AJ, 97, 139

\noindent Lockman, F.J., 1984, ApJ, 283, 90

\noindent Lockman, F.J., \& Gehman, C.S. 1991, ApJ, 382, 182

\noindent Malhotra, S. 1994, ApJ, 433, 687

\noindent Malhotra, S. 1995, ApJ, 448, 138

\noindent McKee, C.F., \& Ostriker, J.P. 1977, ApJ, 218, 148

\noindent Mendez, R.A., \& van Altena, W.F. 1998, A\&A, 330, 910

\noindent Mera, D., Chabrier, G., \& Schaeffer, R. 1998, A\&A, 330, 953

\noindent Narayan, C.A., \& Jog, C.J. 2002, A\&A, 390, L35.

\noindent Olling, R.P. 1995, AJ, 110, 591

\noindent Oort, J.H. 1962, in The Distribution and Motion of Interstellar
Matter in Galaxies, IAU Symp. 15, ed. L. Woltjer (New York: Benjamin), 3

\noindent Porcel, C., Garzon, F., Jimenez-Vicente, J., \& Battaner, E.
1998, A\&A, 330, 136

\noindent Press, W.H., Flannery, B.P., Teukolsky, S.A., \& Vetterling, 
W.T. 1986, Numerical Recipes (Cambridge: Cambridge Univ. Press), chap. 6.

\noindent Rivolo, A.R., Solomon, P.M., \& Sanders, D.B. 1986,
ApJ, 301, L19

\noindent Rohlfs, K. 1977, Lectures on Density Wave Theory,
(Berlin : Springer-Verlag).

\noindent Ruphy, S., et al. 1996, A\&A, 313, L21

\noindent Sanders, D.B., Solomon, P.M., \& Scoville, N.Z. 1984, ApJ,
276, 182
 
\noindent Scoville, N.Z., \& Sanders, D.B. 1987, in Interstellar 
Processes, eds. D.J. Hollenbach \& H.A. Thronson (Dordrecht: Riedel), 21

\noindent Spitzer, L. 1942, ApJ, 95, 329

\noindent Spitzer, L. 1978, Physical Processes in the Interstellar Medium
 (New York: John Wiley)

\noindent Stark, A.A. 1984, ApJ, 281, 624

\noindent van der Kruit, P.C., 1988, A\&A, 192 , 117

\noindent van der Kruit, P.C., \& Searle, L. 1981a, A\&A, 95, 105

\noindent van der Kruit, P.C., \& Searle, L. 1981b, A\&A, 95, 116

\noindent Wouterloot, J.G.A., Brand, J., Burton, W.B., \& Kwee, K.K. 1990,
A\&A, 230, 21

\newpage

\centerline{\bf Figure Legends}

\noindent {\bf FIG. 1a.} $\: - \:$  A plot of HI  vertical
scaleheight versus galactocentric radius.
The joint potential approach gives a theoretical curve (solid line) which is
in a better agreement with observations than the curve obtained
using the stellar potential alone (dashed line), particularly in the 
range of 5-10 kpc.

\noindent {\bf FIG. 1b.} $\: - \:$  A plot of H$_2$ scaleheight
versus radius. 
The scaleheight obtained on using the joint potential (solid line)
agree well with observations over the entire radial range.

\noindent {\bf FIG. 1c.} $\: - \:$  The stellar disk scaleheight
versus radius obtained with the stellar potential alone (dashed line)
 and for the joint potential (solid line). 
The joint potential approach gives a much more moderate flaring and these 
results match with the observational data of Kent et al. (1991).

\noindent {\bf FIG. 2a.} $\: - \:$  The HI scaleheight versus radius
obtained for the joint potential, and where
a linear variation of HI velocity dispersion with a slope of -0.8 
kms$^{-1}$kpc$^{-1}$ has been introduced.
This gives results that are in good agreement with observations in the 
entire radial region studied, and the fit is better than in 
Fig 1a (solid line). 
 
\noindent {\bf FIG. 2b.} $\: - \:$  The vertical scaleheight
versus radius for H$_2$ gas (solid line), obtained using a linearly
varying HI velocity dispersion as in Fig. 2a.
The agreement with observations (crosses) seen here is very close to 
that in Fig. 1b. 
This implies that the introduced variation in the HI velocity dispersion
does not affect the H$_2$ scaleheight noticeably.

\noindent {\bf FIG. 2c.} $\: - \:$  A plot of the predicted stellar 
scaleheight of our Galaxy  versus radius, obtained using a linearly
varying HI velocity dispersion as in Fig. 2a. 
The results are similar to that in Fig. 1c.
Thus the introduced variation in the HI velocity dispersion does not affect
the stellar scaleheight noticeably.

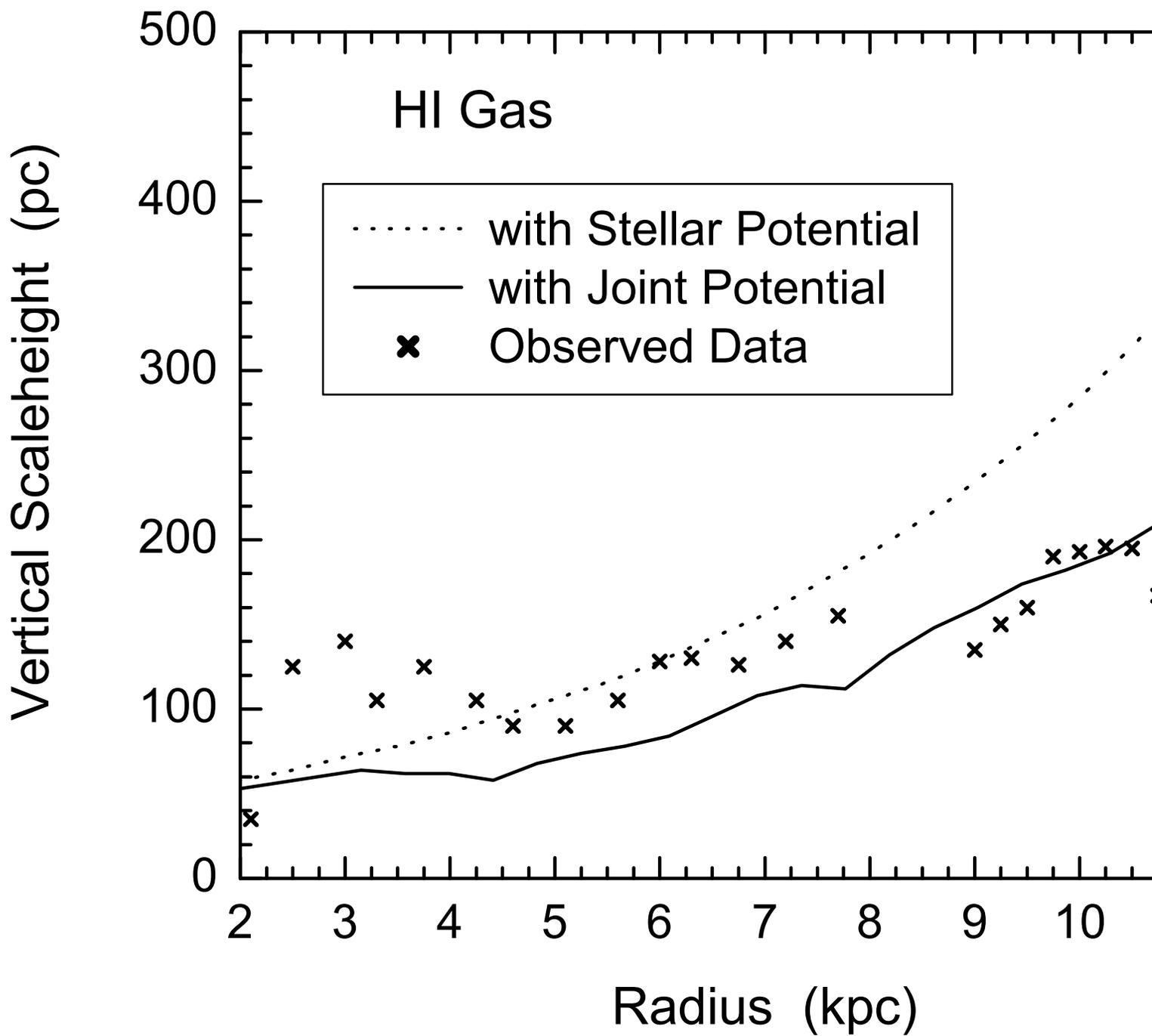
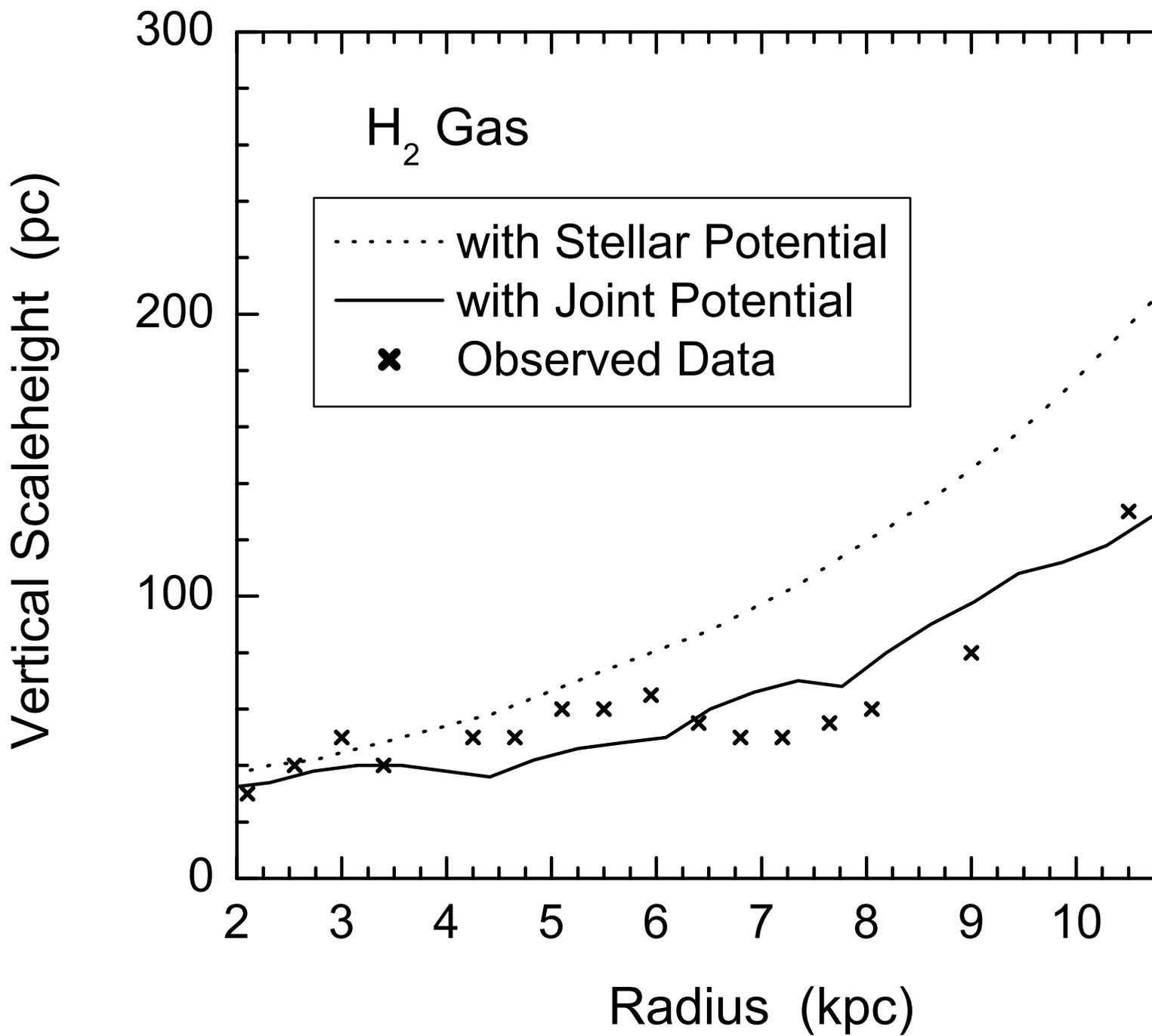
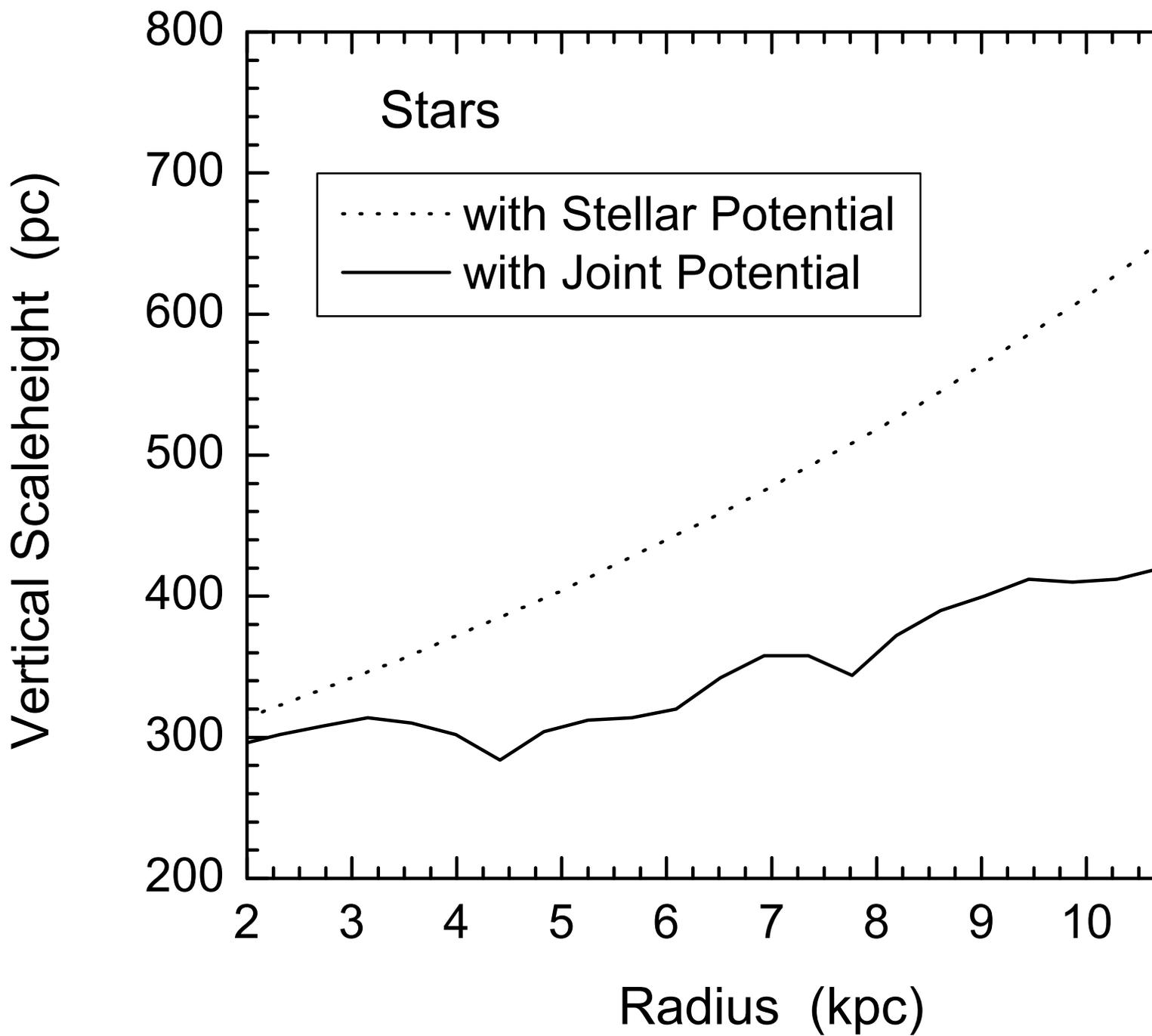
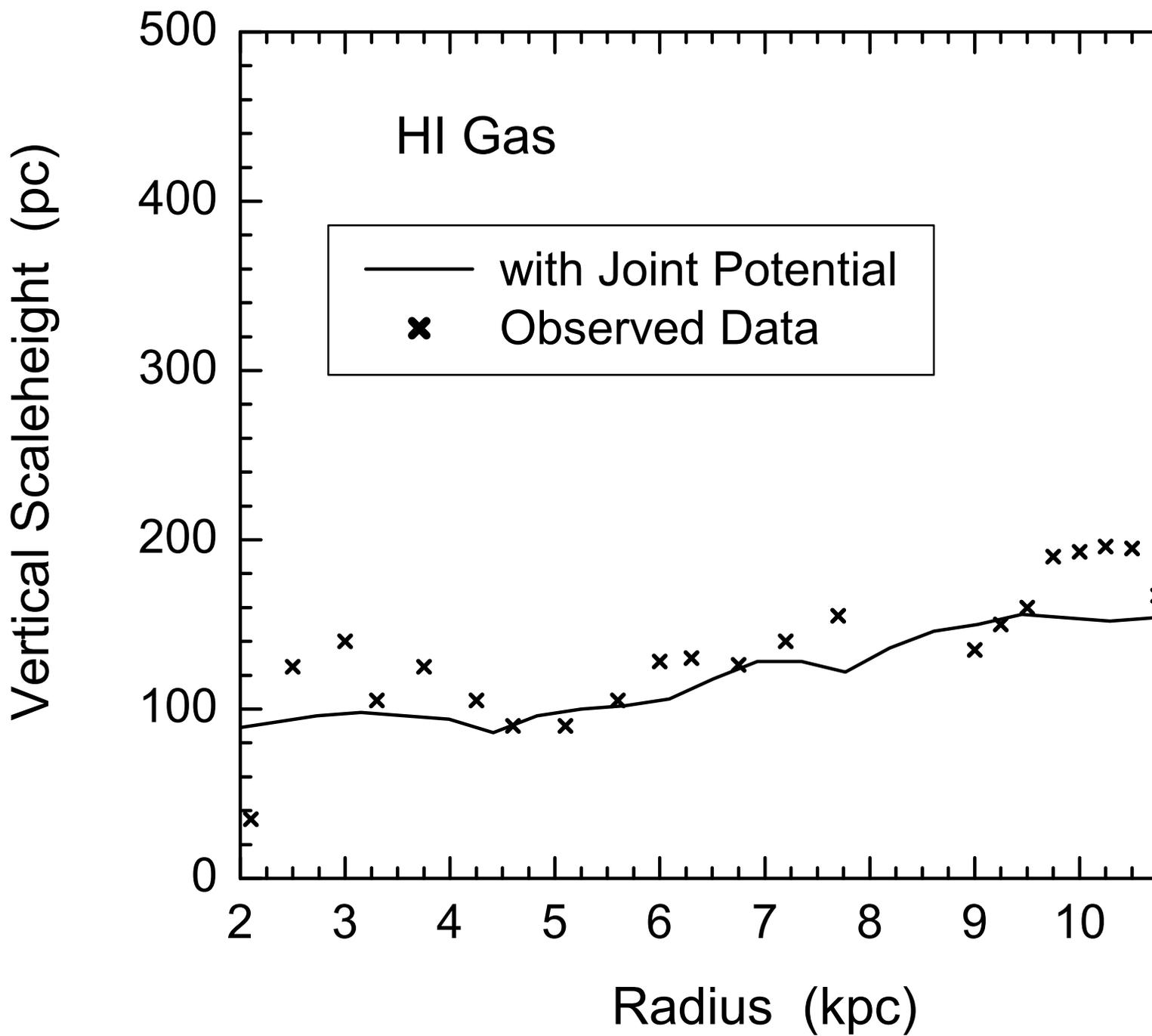
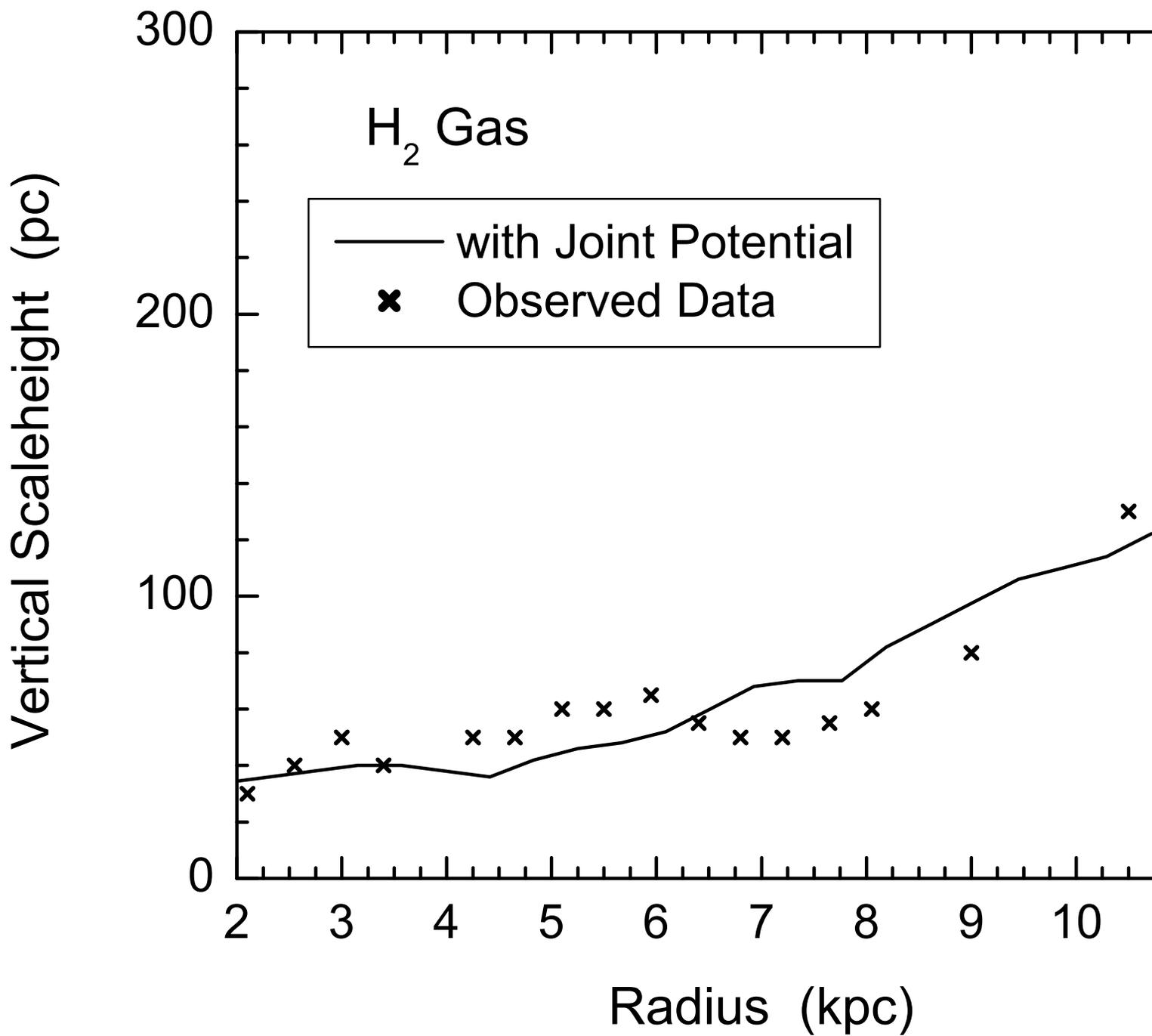
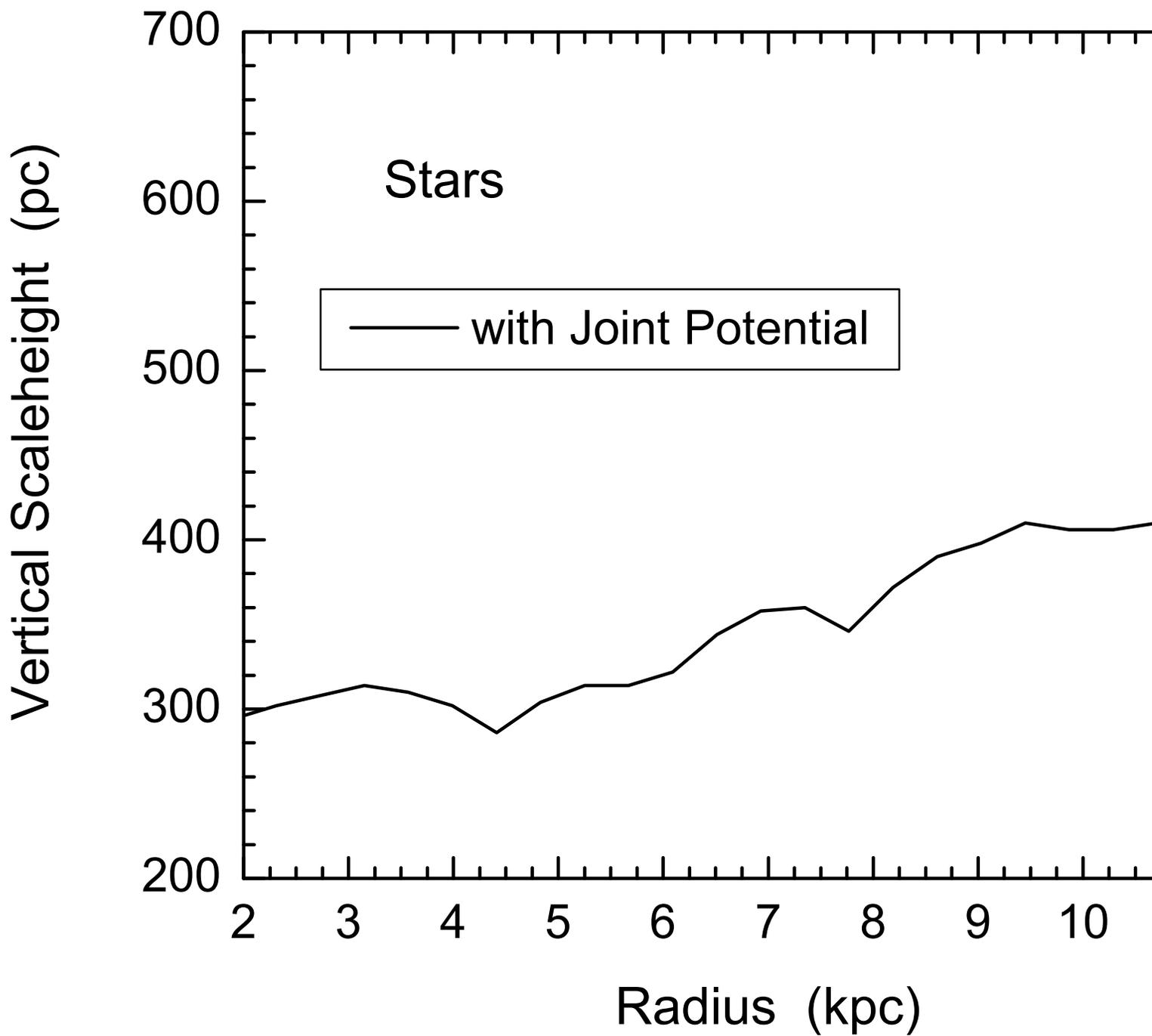
\end{document}